\documentclass[twocolumn, preprintnumbers,prx,aps,amssymb,showpacs,superscriptaddress]{revtex4}
\usepackage{graphicx}
\usepackage{dcolumn}
\usepackage{wasysym}
\usepackage{bm} \usepackage{color}
\usepackage{ulem}
\begin{document}

\newcommand{\SRO}{Sr$_{2}$RuO$_4$}

\newcommand{\TN}{$T_{\rm N}$}
\newcommand{\Tc}{$T_{\rm c}$}

\newcommand{\Ka}{$\kappa_{\text{a}}$}
\newcommand{\Kc}{$\kappa_{\text{c}}$}
\newcommand{\Kzero}{$\kappa_{\text{0}}/T$}
\newcommand{\Kzeroa}{$\kappa_{\text{a0}}/T$}
\newcommand{\Kzeroc}{$\kappa_{\text{c0}}/T$}
\newcommand{\KN}{$\kappa_{\text{N}}/T$}

\newcommand{\ie}{{\it i.e.}}
\newcommand{\eg}{{\it e.g.}}
\newcommand{\etal}{{\it et al.}}
\newcommand{\K}{$\kappa/T$}
\newcommand{\Co}{Ba(Fe$_{1-x}$Co$_x$)$_2$As$_2$}
\newcommand{\unitsmu}{$\mu \text{W}/\text{K}^2\text{cm}$}
\newcommand{\unitsm}{$\text{mW}/\text{K}^2\text{cm}$}
\newcommand{\p}[1]{\left( #1 \right)}
\newcommand{\Dd}[2]{\frac{\text{d} #1}{\text{d}#2}}
\newcommand{\Hc}{$H_{\rm c2}$}

\title{

Vertical line nodes in the superconducting gap structure of \SRO

}


\author{E.~Hassinger}
\email{elena.hassinger@cpfs.mpg.de}
\affiliation{D\'epartement de physique \& RQMP, Universit\'e de Sherbrooke, Sherbrooke, Qu\'ebec J1K 2R1, Canada}
\affiliation{Max Planck Institute for Chemical Physics of Solids, 01187 Dresden, Germany}
\affiliation{Canadian Institute for Advanced Research, Toronto, Ontario M5G 1Z8, Canada}

\author{P.~Bourgeois-Hope} 
\affiliation{D\'epartement de physique \& RQMP, Universit\'e de Sherbrooke, Sherbrooke, Qu\'ebec J1K 2R1, Canada}

\author{H.~Taniguchi}
\affiliation{Department of Physics, Graduate School of Science, Kyoto University, Kyoto 606-8502, Japan}

\author{S.~Ren\'e~de~Cotret} 
\affiliation{D\'epartement de physique \& RQMP, Universit\'e de Sherbrooke, Sherbrooke, Qu\'ebec J1K 2R1, Canada}

\author{G.~Grissonnanche} 
\affiliation{D\'epartement de physique \& RQMP, Universit\'e de Sherbrooke, Sherbrooke, Qu\'ebec J1K 2R1, Canada}

\author{M.~S.~Anwar}
\affiliation{Department of Physics, Graduate School of Science, Kyoto University, Kyoto 606-8502, Japan}

\author{Y.~Maeno}
\affiliation{Canadian Institute for Advanced Research, Toronto, Ontario M5G 1Z8, Canada}
\affiliation{Department of Physics, Graduate School of Science, Kyoto University, Kyoto 606-8502, Japan} 

\author{N.~Doiron-Leyraud} 
\affiliation{D\'epartement de physique \& RQMP, Universit\'e de Sherbrooke, Sherbrooke, Qu\'ebec J1K 2R1, Canada}

\author{Louis~Taillefer}
\email{louis.taillefer@usherbrooke.ca}
\affiliation{D\'epartement de physique \& RQMP, Universit\'e de Sherbrooke, Sherbrooke, Qu\'ebec J1K 2R1, Canada}
\affiliation{Canadian Institute for Advanced Research, Toronto, Ontario M5G 1Z8, Canada}

\date{\today}


\begin{abstract}

There is strong experimental evidence that the superconductor \SRO~has a chiral {\it p}-wave order parameter. 
This symmetry does not require that the associated gap has nodes, yet specific heat, ultrasound and thermal conductivity measurements 
 indicate the presence of nodes in the superconducting gap structure of \SRO. 
Theoretical scenarios have been proposed to account for the existence of 
 deep minima
or 
accidental nodes (minima tuned to zero or below by material parameters)
within a $p$-wave state. 
Other scenarios propose chiral $d$-wave and $f$-wave states, with horizontal and vertical line nodes, respectively.
To elucidate the nodal structure of the gap,
it is essential to know whether the lines of nodes (or minima)
are vertical (parallel to the tetragonal $c$ axis) or horizontal (perpendicular to the $c$ axis). 
Here, we report thermal conductivity measurements on single crystals of \SRO~down to 50 mK for currents parallel and perpendicular to the {\it c} axis.
We find that there is substantial quasiparticle transport in the $T = 0$ limit for both current directions.
A magnetic field $H$ 
immediately
excites quasiparticles with velocities both in the basal plane and in the $c$ direction.
Our data down to \Tc/30~and down to \Hc/100 show no evidence that the nodes are in fact deep minima.
Relative to the normal state, the thermal conductivity of the superconducting state  
is found to be
very similar
 for the two current directions, from $H=0$ to $H=$~\Hc.
These findings show that 
the gap structure of \SRO~consists of vertical line nodes.
%
This rules out a chiral $d$-wave state.
%
%
Given that the $c$-axis dispersion (warping) of the Fermi surface in \SRO~varies strongly from surface to surface,
the
small
$a-c$ anisotropy 
suggests that the line nodes are
present on all three sheets of the Fermi surface.
%
If imposed by symmetry, vertical line nodes would 
be inconsistent with a $p$-wave order parameter for \SRO.
%
To reconcile the gap structure revealed by our data with a $p$-wave state, a mechanism
must be found that produces accidental line nodes in \SRO.


\end{abstract}


\maketitle

\section{INTRODUCTION}

\SRO~is one of the rare materials in which {\it p}-wave superconductivity is thought to be realized. 
Nuclear magnetic resonance
\citep{Ishida_1998_Nature, Ishida_2016_PRB} and neutron scattering \citep{Duffy_2000_PRL} measurements find no
drop
in the spin susceptibility below the superconducting transition temperature \Tc,
strong evidence in favour of spin-triplet pairing.
Measurements of muon spin rotation \citep{Luke_1998_Nature, Luke_2000_PhysicaB} and the polar Kerr angle \citep{Xia_Kerr} 
show that time-reversal symmetry is spontaneously broken below \Tc. 
These results (and others) have led to the view that  \SRO~has a chiral {\it p}-wave order parameter, 
with a $d$-vector given by ${\mathbf d} = \Delta_0{\mathbf z}(k_x\pm ik_y)$
\citep{Mackenzie_2003_RMP,Maeno_2012_JPSJ,Kallin_2012_RPP}.
Nevertheless, the symmetry of the superconducting order parameter in \SRO~is still under debate
\citep{,Maeno_2012_JPSJ,Kallin_2012_RPP}.
One of the problems is that
although the gap structure of a chiral {\it p}-wave order parameter
is not required by symmetry to go to zero, {\it i.e.} to have nodes, anywhere on a two-dimensional Fermi surface,
there are in fact low-energy excitations deep inside the superconducting state of \SRO, as detected in 
the specific heat \citep{Nishizaki_1999_JLTP,Nishizaki_2000_JPSJ,Deguchi_2004_PRL,Deguchi_2004_JPSJ}, ultrasound attenuation \citep{Lupien_2001_PRL} 
and penetration depth \citep{Bonalde_2000_PRL} at very low temperature.
%
%
Theoretical scenarios have been proposed to account for those excitations in terms of 
either accidental nodes that are perpendicular to the tetragonal $c$ axis (\ie~`horizontal') \citep{Rice_Hnodes} 
or deep minima in the superconducting gap along lines parallel to the $c$ axis (\ie~`vertical')
\citep{Nomura_2005_JPSJ, Raghu_2010_PRL, Thomale_2013_EPL, Scaffidi_2014_PRB}.
The latter vary in depth from sheet to sheet on the three-sheet Fermi surface of \SRO.
On the large $\gamma$~sheet, the gap develops deep minima in the $a$ direction because an odd-parity order parameter must go to zero at the zone boundary.

These scenarios are difficult to reconcile with the 
specific heat and thermal conductivity of \SRO.
%
When plotted as $C_e/T$ vs $T$, the electronic specific heat $C_e$ of \SRO~is perfectly linear
below $\sim$~\Tc/2, down to the lowest temperature \citep{Nishizaki_1999_JLTP,Nishizaki_2000_JPSJ,Deguchi_2004_PRL,Deguchi_2004_JPSJ}.
%
%
Gap minima of various depths inevitably lead to deviations from perfect linearity
in $C_e/T$ vs $T$ \cite{Nomura_2005_JPSJ}.
In the clean limit,
a truly linear behaviour can only be obtained if the minima on all three sheets are so deep that they extend to negative values,
thereby producing accidental nodes.

\begin{figure}[t]
\centering
\includegraphics[width=8.6cm]{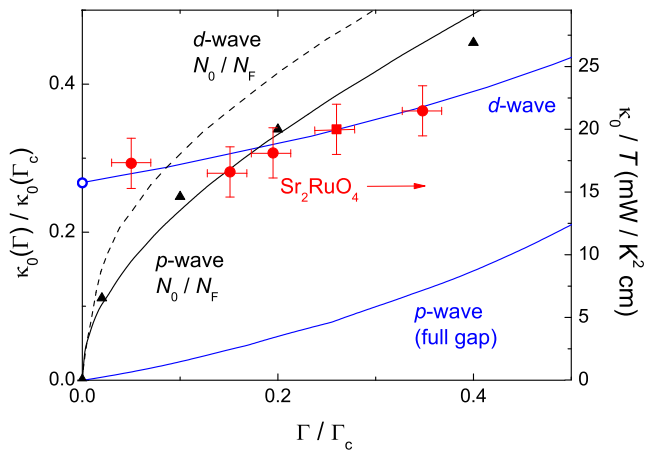}
\caption{
Residual linear term in the thermal conductivity, \Kzero, as a function of impurity scattering rate $\Gamma$,
both normalized to unity at $\Gamma = \Gamma_{\rm c}$, the critical scattering rate needed to suppress superconductivity.
The blue lines are theoretical calculations for a $d$-wave state \cite{Maki_1995_EPL} and a fully-gapped $p$-wave state \cite{Maki_1999_EPL}, as indicated (left axis).
In the clean limit ($\Gamma \to 0$), \Kzero~vanishes in the $p$-wave case while it reaches a non-zero value in the $d$-wave case (open circle),
whose value is given by Eq.~1, estimated at \Kzero~$= 15.8$~mW / K$^2$ cm in \SRO~(see text).
The experimental values of \Kzero~measured in \SRO~are plotted as red symbols (right axis; circles, \cite{Suzuki_2002_PRL}; square, this work),
taking $\hbar \Gamma_{\rm c} = k_{\rm B} T_{\rm c0}$.
The black lines are the zero-energy density of states $N_0$, normalized by the normal-state value $N_{\rm F}$ 
(left axis; solid, full-gap $p$-wave \cite{Maki_2000_EPL}; dashed, $d$-wave \cite{Maki_1995_PRB}).
Black triangles show $N_0 / N_{\rm F}$ for a $p$-wave state with a deep gap minimum ($\Delta_{\rm min} \simeq \Delta_{\rm max} / 4$) \cite{Miyake_1999_PRL}.
}
\label{Fig0}
\end{figure}

The in-plane thermal conductivity \Ka$(T)$ of \SRO~decreases smoothly 
down to the lowest measured temperature,
and it extrapolates to a large residual linear term, \Kzero, at $T=0$ \citep{Suzuki_2002_PRL}.
%
This residual linear term 
is robust against impurity scattering,
and virtually unaffected by a 10-fold increase in scattering rate \citep{Suzuki_2002_PRL}.
This is the classic behaviour of a nodal superconductor whose nodes are imposed by symmetry
\citep{Maki_1995_EPL,Maki_1995_PRB,Graf_1996_PRB,Durst_2000_PRB} 
(Fig.~\ref{Fig0}),
 as in the $d$-wave state of cuprate superconductors \citep{Taillefer_1997_PRL}.
It comes from the linear energy dependence of the density of states at low energy,
which produces a compensation between the growth in the density of quasiparticles
and the decrease in their mean free path as a function of impurity scattering \citep{Graf_1996_PRB}.
%
Such a compensation does not occur in a nodeless $p$-wave state \citep{Maki_1999_EPL,Maki_2000_EPL}
(Fig.~\ref{Fig0}),
nor does it occur
for accidental nodes in an $s$-wave state \citep{Mishra_2009_PRB}.
%

In summary, the known properties of \K~and $C_e$ in \SRO~strongly suggest that the low-energy quasiparticles 
in the superconducting state come from nodes in the gap, not from deep minima.
Because accidental nodes do not occur naturally in the chiral $p$-wave state that is widely proposed for \SRO,
it is important to establish the presence of nodes.
Moreover, because other proposed states have symmetry-imposed line nodes that are either 
horizontal (chiral $d$-wave state \cite{Mazin_2005_PRL})
or vertical ($f$-wave state \cite{Hasegawa_2005_JPSJ,Graf_2000_PRB}),
we need to determine whether line nodes are vertical or horizontal.


%
Existing experimental evidence
 on the direction of line nodes in \SRO~is contradictory. 
Measurements of the heat capacity as a function of the angle made by a magnetic field $H$ applied in the basal plane (normal to the $c$ axis)
relative to the $a$ axis ([100] direction) reveal a small four-fold variation below 0.25~K that is consistent with
vertical line nodes along the $\Gamma$M directions \citep{Deguchi_2004_PRL,Deguchi_2004_JPSJ}.
However, no such angular variation was detected in the heat conduction down to 0.3~K \citep{Izawa_2001_PRL, Tanatar_2001_PRL, Tanatar_2001_PRB}.
Moreover, ultrasound attenuation in \SRO~is rather isotropic in the plane,
unexpected if line nodes are vertical \citep{Lupien_2001_PRL}.

 In this Letter, 
we shed new light on the gap structure of \SRO~by using the directional power of thermal conductivity
to determine whether the line nodes are vertical or horizontal.
 %
 In particular, we probe nodal quasiparticle motion along the $c$ axis as $T \to 0$, 
 from measurements of \Kc, the thermal conductivity along the $c$ axis, down to \Tc /30 (50~mK).
 We observe a substantial residual term \Kzero~in the $c$ direction at $H=0$.
 Moreover, \Kzero~is rapidly enhanced by a magnetic field, 
even as low as \Hc~/ 100.
 This 
 confirms
 that the line nodes in \SRO~are not deep minima
 and it 
shows they must be vertical.
Furthermore, quantitative analysis 
suggests that the line nodes are present on all three Fermi surfaces.
If the vertical line nodes are imposed by symmetry, then, by virtue of Blount's theorem \cite{Blount},
they would rule out a spin-triplet state, such as the proposed $p$-wave state \cite{Kobayashi}.
Conversely, if \SRO~is indeed a $p$-wave superconductor, then a reason must be found for the presence
of 
accidental line nodes in its gap function.
%
Note that the obvious spin-singlet state that breaks time-reversal symmetry has symmetry-imposed line nodes
that are horizontal, not vertical \cite{Mazin_2005_PRL}.


\section{METHODS}

Single crystals of \SRO~were grown by the floating-zone method~\citep{Mao_2000}
and annealed in oxygen flow at 1080~$^{\circ}$C for 8 days.
Both samples were cut 
into rectangular platelets
from the same crystal rod 
that contained very few Ru inclusions ($\sim 3$~inclusions / mm$^2$).
No $3\,$K anomaly was detected in either the susceptibility of the large annealed crystal or the resistivity of the small measured samples.
The {\it a}-axis sample had a length of 4.0~mm along the $a$~axis, and a cross-section of $0.3 \times 0.18$~mm$^2$.
The {\it c}-axis sample had a length of 1.0~mm along the $c$~axis, and a cross-section of $0.4 \times 0.42$~mm$^2$.
The geometric factor of the {\it a}-axis sample was refined
by normalizing the room-temperature resistivity to the well-established literature value of 
$\rho_{\rm a}(300~{\rm K}) = 121~\mu\Omega$~cm \citep{Mackenzie_PRL_1998}. 
The geometric factor of the {\it c}-axis sample was
calculated
from sample dimensions and contact separation. 
The value we find is $\rho_{\rm c}(300~{\rm K}) = 33$~m$\Omega$~cm, in the range of reported values \citep{Tyler_1998_PRB,Maeno_1994_Nature}. 
Contacts were made with silver epoxy (Epo-Tek H20E)  heated at $450\,^{\circ}\mathrm{C}$ for 1 hour in oxygen flow. 
Silver wires were then glued on with silver paint.
From our thermal conductivity measurements, we obtain a superconducting transition temperature $T_c = 1.2$~K,
consistent with the measured residual resistivity of our $a$-axis~sample, $\rho_{\rm a0} = 0.24~\mu\Omega$~cm~\citep{Mackenzie_PRL_1998}.
%
The thermal conductivity was measured using a one heater-two thermometer method \cite{ReidPRB2010}, 
with an applied temperature gradient  of 2-5\%
of the sample temperature. 
%
Measurements where carried out for two directions of the magnetic field $H$:
$H \parallel a$ and $H \parallel c$.
For $H \parallel a$, the field was aligned to within $1^{\circ}$ of the $a$ axis, and perpendicular to the heat current.
(For this field direction, a misalignment of  $1^{\circ}$ can cause a decrease of \Hc~by 0.1~T \citep{Kittaka_2009}.)
%
The field was always changed at $T >$~\Tc.


\section{H = 0 : IN-PLANE TRANSPORT}

\begin{figure}[t]
\centering
\includegraphics[width=7.8cm]{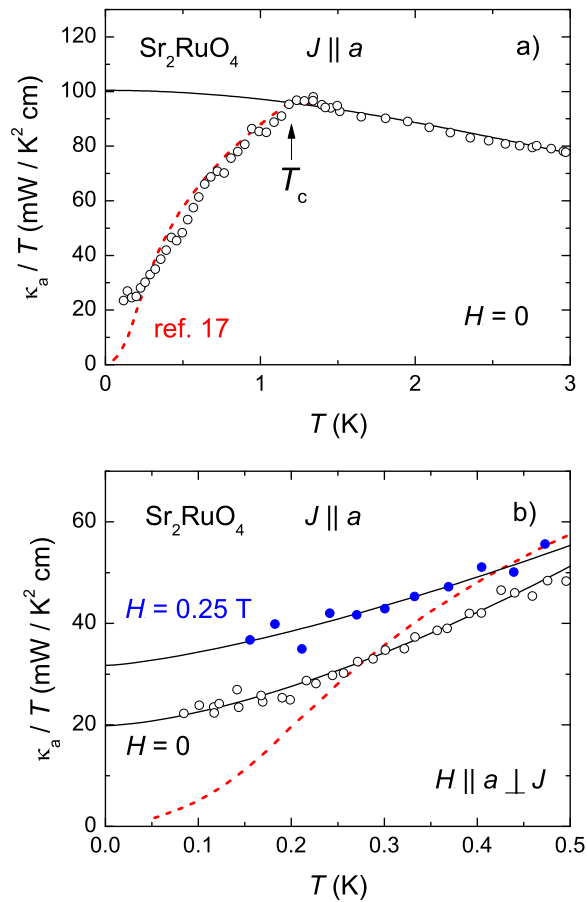}
\caption{
a) 
In-plane ($a$-axis) thermal conductivity \Ka($T$)~of \SRO~at $H = 0$ 
(open circles). 
The black line is a Fermi-liquid fit to the normal-state data, $\kappa_{\rm N}/T = L_0 / (a + bT^2)$, extended below \Tc.
The arrow marks the location of the superconducting transition temperature, \Tc~=~1.2~K,
defined as the temperature below which $\kappa/T$ deviates from its normal-state behavior.
Note that the contribution of phonons to \Ka, $\kappa_{\rm p}$, is negligible up to 3~K, 
so that \Ka~$\simeq$~$\kappa_{\rm e}$, the electronic contribution.
The red dashed line is a calculation for a three-band model of a $p$-wave state with 
deep minima
in the gap structure 
\cite{Nomura_2005_JPSJ} (see text).
It provides a good description of the data at high temperature, but it fails below 0.3~K.
b)
Zoom at low temperature.
Data taken in a magnetic field $H = 0.25$~T ($H \parallel a$) are also shown (blue dots).
The solid black
lines are a fit of the data to the form $\kappa/T = \kappa_0/T + c T^n$. 
}
\label{Fig1}
\end{figure}

Fig.~\ref{Fig1}a shows the thermal conductivity of \SRO~in zero field, for the current in the plane 
($J  \parallel a$).
%
%
The conductivity \Ka~is completely dominated by the electronic contribution \citep{Suzuki_2002_PRL}, $\kappa_{\rm e}$, 
so that $\kappa_{\rm e} >> \kappa_{\rm p}$ up to
$\sim 3$~K, 
where $\kappa_{\rm p}$ is the phonon conductivity. 
In Fig.~\ref{Fig1}a,  a Fermi-liquid fit to the normal-state data (above \Tc) yields
\KN~=~$L_0 / (a + bT^2)$, where $L_0 \equiv (\pi^2/3) (k_{\mathrm B}/e)^2$, with $a = 0.24~\mu\Omega$~cm and $b = 8~{\rm n}\Omega$~cm/K$^2$. 
We see that the Wiedemann-Franz law is satisfied, with $a = \rho_{a0}$.
%

Fig.~\ref{Fig1}b shows a zoom of the data at low temperature,
seen to extrapolate to \Kzero~=~$20 \pm 2$ \unitsm, a large residual linear term in excellent agreement 
with the value reported for \SRO~samples of similar \Tc~\citep{Suzuki_2002_PRL}
(Fig.~\ref{Fig0}).
%
In the limit of a vanishing impurity scattering rate $\Gamma$, 
whence
\Tc~$\to 1.5$~K, \Kzero~=~$17 \pm 2$ \unitsm~\citep{Suzuki_2002_PRL}.
%
%
A ten-fold increase in $\Gamma$ only yields a modest increase in \Kzero~(Fig.~\ref{Fig0}).
%
Such a weak dependence of \Kzero~on $\Gamma$ is precisely the behavior expected of a superconductor with symmetry-imposed line nodes,
whereby the impurity-induced growth in the quasiparticle density of states is compensated by a corresponding decrease in mean free path,
as in a $d$-wave superconductor \citep{Maki_1995_EPL,Graf_1996_PRB,Durst_2000_PRB}
(Fig.~\ref{Fig0}).
%
(By contrast, accidental nodes and deep minima
in an $s$-wave superconductor
are not robust against impurity scattering, 
so they will in general
be lifted or be made shallower, respectively, causing \Kzero~to vanish or decrease with impurity concentration, respectively \citep{Mishra_2009_PRB}.)
%
In other words, the remarkable fact that \Kzero~remains~large \citep{Suzuki_2002_PRL} 
even when the zero-energy density of states vanishes \citep{Nishizaki_1999_JLTP} 
as $\Gamma \to 0$ in \SRO~is the clear signature of a line node.
Indeed, while impurities in a $p$-wave superconductor without nodes do induce a zero-energy density of states \citep{Maki_1999_EPL,Maki_2000_EPL,Miyake_1999_PRL},
 the associated \Kzero~vanishes as $\Gamma \to 0$ \citep{Maki_1999_EPL,Maki_2000_EPL}
(Fig.~\ref{Fig0}) because the impurity-induced states are localized.

The  
magnitude
of \Kzero~at $\Gamma \to 0$ can be evaluated theoretically from a knowledge of the Fermi velocity
$v_{\rm F}$ and the gap velocity at the node, $v_{\Delta}$.
For a $d$-wave gap on a single 2D Fermi surface \citep{Durst_2000_PRB} :
\begin{equation}
\frac{\kappa_{\text{0}}}{T} = \frac{k_{\rm B}^2}{3\hbar}\frac{1}{c} \left(\frac{v_{\rm F}}{v_{\Delta}} + \frac{v_{\Delta}}{v_{\rm F}}\right)~,
\end{equation}
where $c$ is the interlayer separation along the $c$ axis and $v_{\Delta} = 2 \Delta_0 / \hbar k_{\rm F}$, in terms of the Fermi wavevector $k_{\rm F}$
and the gap maximum $ \Delta_0$, in $\Delta({\rm \phi}) = \Delta_0 \cos{\rm 2\phi}$.
This expression works very well for overdoped cuprate superconductors such as YBa$_2$Cu$_3$O$_7$ and Tl$_2$Ba$_2$CuO$_{6 + \delta}$ \citep{Reid_2012_SUST}, 
quasi-2D metals where the pairing symmetry 
is established to be $d$-wave.
Let us use Eq.~1 to estimate \Kzero~in \SRO.


The Fermi surface of \SRO~is quasi two-dimensional and it has been characterized experimentally 
in exquisite detail~\citep{Bergemann_2003_AP}. 
It consists of three cylinders: 
two at the center of the Brillouin zone ($\beta$ and $\gamma$) and one ($\alpha$) at the corner. 
The values of $k_{\rm F}$ and $v_{\rm F}$ are known precisely for each.
In a $d_{x^2-y^2}$-wave state, 
each cylinder
would have four vertical line nodes 
(along the $x = \pm~y$ directions).
Assuming the same gap on each Fermi surface and using the weak-coupling expression $\Delta_0 = 2.14~k_{\rm B}$\Tc, we get 
\Kzero~=~3.7, 7.3 and 4.8~\unitsm~for the $\alpha$, $\beta$ and $\gamma$ sheets, respectively, giving a total conductivity
\Kzero~=~15.8~\unitsm~
(open circle on the $y$ axis of Fig.~\ref{Fig0}).
%
%
This theoretical value is in remarkably good agreement with the measured value $\kappa_{\rm a0}/T = 17 \pm 2$ \unitsm~\citep{Suzuki_2002_PRL},
consistent with
line nodes on all three Fermi surfaces.
%
%
%

One may ask whether our data are compatible with deep minima instead of nodes.
In Fig.~\ref{Fig1}, we compare our data with calculations for a model of \SRO~in the clean limit
where the gap has 
symmetry-related minima along the $a$ axis on the $\gamma$ sheet and
very deep minima 
along the zone diagonals
on the $\alpha$~and $\beta$~sheets,
that result
 from
the model interaction  \cite{Nomura_2005_JPSJ}.
The deepest minima are on the $\beta$ sheet, where the gap goes down to a value 30 times
smaller than its maximal value 
(on the $\gamma$ sheet).
%
We see that while the model works well for $T >$~\Tc/4, it 
fails
at lower $T$, forced as it is to go to zero at $T \to 0$ since the 
gap does not have nodes.
This comparison shows that our data are inconsistent 
even with minima so deep that $\Delta_{\rm min} \simeq \Delta_{\rm max}/30$.
Taking into account the perfectly linear $T$ dependence of the specific heat below $\sim$~\Tc/2,
the case against deep minima in the gap is compelling, for the combined data
require that $\Delta_{\rm min} \simeq \Delta_{\rm max}/100$ on each of the three Fermi surfaces -- a rather 
artificial situation.
Note that adding impurities to the calculation by Nomura \cite{Nomura_2005_JPSJ} would produce 
a non-zero \Kzero, thereby achieving better agreement with experiment (Fig.~\ref{Fig1}).
However, that magnitude of \Kzero~would be expected to decrease rapidly as $\Gamma_0 \to 0$ \cite{Maki_1999_EPL,Maki_2000_EPL},
contrary to what is observed experimentally \cite{Suzuki_2002_PRL} 
(Fig.~\ref{Fig0}).

%
%


\begin{figure}[t]
\centering
\includegraphics[width=8.8cm]{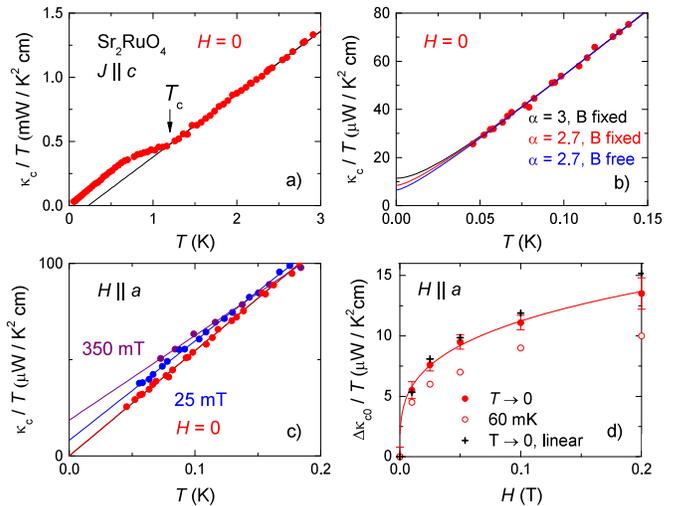}
\caption{
Out-of-plane ($c$-axis) thermal conductivity of \SRO.
a)
At $H = 0$. 
The black line is a linear fit to the normal-state data. 
The arrow marks the location of \Tc~=~1.2\,K.
In this direction, 
$\kappa_{\rm e} << \kappa_{\rm p}$, 
so that the purely electronic term is obtained as \Kc($T$)/$T$ in the $T=0$ limit.
b)
Zoom on the data at low temperature, at $H = 0$ (red dots).
The black line is a fit of the data to Eq.~3 below 0.35~K,
with the phonon conductivity in the $T \to 0$ limit given by
$\kappa_{\rm p} = B T^{\alpha}$, with $\alpha = 3.0$ and $B$ given by
sound velocity and sample dimensions (see text).
The other two lines are the same fit but with $\alpha = 2.7$, to take into account 
the effect of specular reflection, and $B$ whether fixed (red line) or free (blue line) (see text).
c)
Same data as in b) (red, $H=0$), compared with data in a magnetic field 
$H = 25$~mT (blue) and $H = 0.35$~T (burgundy), with $H \parallel a$.
d)
Increase in $\kappa_{\rm c}/T$ with field $H \parallel a$, where $\kappa_{\rm c}/T$ is either measured at $T = 60$~mK (open circles) 
or extrapolated to $T=0$, whether linearly as in panel (c) (crosses) or through a fit as in panel (b), red line (full red dots). The red line is a guide to the eye.}
\label{Fig2}
\end{figure}



\section{H = 0 : C-AXIS TRANSPORT}

Fig.~\ref{Fig2}a shows the conductivity out of the plane, \Kc$(T)$ ($J  \parallel c$). It is completely dominated by the phonon contribution $\kappa_{\rm p}$, since in this direction
$\kappa_{\rm e}$ is some 2000 times smaller than in the plane (estimated from the resistivity anisotropy).
Because of this, the only way to extract the electronic contribution of interest is to obtain the purely fermionic residual linear term
at $T=0$.
A zoom on the $c$-axis conductivity at low temperature is shown in 
Figs.~\ref{Fig2}b and~\ref{Fig2}c.
We see that \Kc$/T$ is linear below 0.2~K. 
We attribute this linear behavior of $\kappa_{\rm p}/T$,
also observed in overdoped cuprate superconductors \citep{Hawthorn_2007_PRL},
to the scattering of phonons by nodal quasiparticles,
as discussed theoretically in ref.~\onlinecite{Smith_2005_PRB}.

A linear fit to $\kappa_{\rm{c}}/T$ extrapolates to $\kappa_{\rm{c0}}/T = 0.0 \pm 3$\,\unitsmu
~(Fig.~\ref{Fig2}c).
However, \Kc$(T)/T$ cannot continue linearly all the way down to $T = 0$,
for this would imply a divergent phonon mean free path, since $l_{\rm p} \propto \kappa/T^3$.
%
The sample boundaries impose an upper bound on $l_{\rm p}$.
For diffuse (non-specular) scattering,
$l_0 = 2\sqrt{S/\pi}$, where $S$ is the sample cross-section normal to the heat flow.
In the ballistic regime at low temperature, where phonons 
are scattered by the (rough) sample boundaries, we have
\citep{Li2008}:
\begin{equation}
\kappa_{\rm p} = \frac{1}{3} C_{\rm p} v_{\rm p} l_{\rm 0} = BT^3~~,
\end{equation}
where $C_{\rm p} = (2 \pi^2 k_{\rm B} / 5) (k_{\rm B} T / \hbar v_{\rm p})^3$ 
is the 
phonon specific heat \citep{Ashcroft} and
$v_{\rm p}$ is the average 
sound velocity.
%
%
%
%
%
$v_{\rm p}$~can be extracted from the 
measured
phonon specific heat $C_{\rm p} / T^3 = 0.197$~mJ/K$^4$ mole = 3.44~J/K$^4$ m \citep{Deguchi_2004_PRL},
giving $v_{\rm p} = 3 284$~m/s, 
a value which is consistent with the 
measured sound velocities in \SRO~\citep{Lupien_2001_PRL}. 
Using Eq.~2, with $l_0 = 0.46$~mm, we get 
$B = 17.3$~mW~/~K$^4$ cm. 

The total thermal conductivity is given by $\kappa/T = \kappa_{\rm c0}/T + \kappa_{\rm p}/T$,
where the first term is electronic and the second term is phononic.
At low $T$, two mechanisms scatter phonons: the sample boundaries, already mentioned, 
and quasiparticles. 
In Eq.~2, the phonon mean free path $l_{\rm 0}$ is replaced by $l_{\rm p} = [1 / l_{\rm 0} + 1 / l_{\rm e}]^{-1}$,
where $l_{\rm e}$ is the mean free path due to electron-phonon scattering,
with $1 / l_{\rm e} \propto T$ \cite{Smith_2005_PRB}.
Therefore, in the regime where the latter process dominates, we get 
$\kappa_{\rm p} = A T^2$, as seen in our data at $T > 0.05$~K 
(Fig.~\ref{Fig2}c).
In the limit $T \to 0$, we expect $\kappa_{\rm p} = B T^3$.
We can therefore fit our data to: 
\begin{equation} 
\kappa / T = \kappa_{\rm c0}/T + B T^2 / (1 + BT/A)~~.
\end{equation} 
Given that $B$ is known and $A$ is fixed by the
slope of $\kappa / T$ above 50~mK,
the only free parameter in the fit is 
the residual linear term $\kappa_{\rm c0}/T$,
due to quasiparticle transport.
A fit to the zero-field data of Fig.~\ref{Fig2}b yields 
$\kappa_{\rm c0}/T = 12 \pm 5~\mu$W~/~K$^2$~cm
(black line).
%
%


Although the cut side surfaces of our $c$-axis sample are rougher than the mirror-like cleaved or as-grown
surface of crystals, there can still be some degree of specular reflection.
This was studied on crystals of the cuprate insulator Nd$_2$CuO$_4$, with sample surfaces roughened by sanding \cite{Li2008}.
At $0.15 <ÊT < 0.3$~K,
$\kappa_{\rm p} = B T^{3}$, with the prefactor $B$ correctly given by the sound velocities and sample dimensions (Eq.~2).
At $T <Ê0.15$~K, specular reflection becomes important and
$\kappa_{\rm p} = B' T^{2.68}$, with $B' = 0.6 B$.
Using the same power law to fit our \SRO~$c$-axis data, namely
$\kappa_{\rm p} = B' T^{\alpha}$
with $\alpha = 2.7$ and $B' = 0.6~B = 10~\mu$W / K$^{3.7}$ cm,
we get the red line in Fig.~\ref{Fig2}b, with
$\kappa_{\rm c0}/T = 8.5~\mu$W~/~K$^2$~cm.
Leaving $B'$ as a free fit parameter yields $\kappa_{\rm c0}/T = 6.5~\mu$W~/~K$^2$~cm (blue line).
We arrive at a value for the residual linear term of 
$\kappa_{\rm c0}/T = 10 \pm 5~\mu$W~/~K$^2$~cm.

\begin{figure}[t]
\centering
\includegraphics[width=8.8cm]{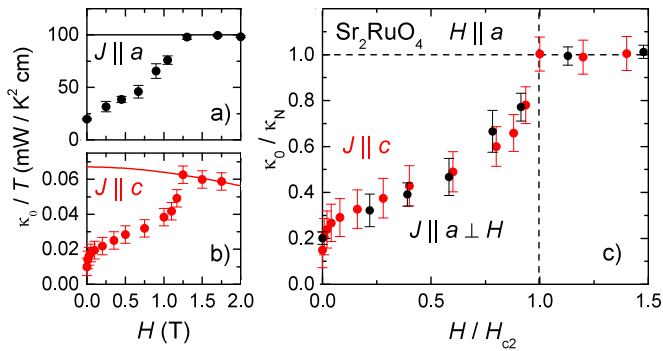}
\caption{
Residual linear term \Kzero~as a function of a magnetic field $H$ applied along the $a$ axis ($H \parallel a$).
a) 
For a current in the plane ($J \parallel a$).
The data points (black dots) are \Kzeroa~obtained by fitting \Ka$/T$~$vs$ $T$ as in Fig.~1b.
The black line is a constant fit to the data above \Hc~
(negligible magnetoresistance for that current direction).
It defines $\kappa_{\text{aN}}/T$~$vs$ $H$, and it is consistent with the $H=0$ value obtained
by 
extrapolating \KN~above \Tc~to $T \to 0$ (Fig.~\ref{Fig1}a).
b) 
Same as in a), but for a current along the $c$ axis ($J \parallel c$).
The data points (red dots) are \Kzeroc~obtained by fitting \Kc$/T$~$vs$ $T$ as in Fig.~\ref{Fig2}b.
Above \Hc, \Kzeroc~decreases slightly due to magnetoresistance
(see text). 
The red line is a fit of the data above \Hc~to \KN~$= a / (b + c H^{2})$,
which defines $\kappa_{\text{cN}}/T$~$vs$ $H$ for this current direction.
The value at $H \to 0$ is $\kappa_{\text{cN}}/T = 67 \pm 7$\,\unitsmu.
c)
Field dependence of \Kzeroa~(black dots) and \Kzeroc~(red dots) normalized to their normal-state value, 
both plotted as (\Kzero)/(\KN)~$vs$ $H/$\Hc, with \Hc~$= 1.25$~T.
For simplicity, we define $\kappa_0 / \kappa_{\rm N} \equiv$~(\Kzero)/(\KN).
The error bars on $\kappa_0 / \kappa_{\rm N}$ come from the combined uncertainties in extrapolating
$\kappa/T$ to $T=0$ to obtain \Kzero~and in extending \KN~below \Hc.
}
\label{Fig3}
\end{figure}

Nodal quasiparticles in \SRO~must therefore have a non-zero $c$-axis velocity.
This rules out horizontal line nodes -- at least in high-symmetry planes (\eg~$k_z = 0$) -- 
and it points immediately to vertical line nodes.
What magnitude of $\kappa_{\rm c0}/T$ do we expect if the line nodes responsible
for the large in-plane $\kappa_{\rm a0}/T$ are vertical?
Assuming all three Fermi surfaces have line nodes, as would be the case for a $d_{x^2-y^2}$~symmetry, 
then the $a$-$c$ anisotropy of nodal quasiparticle transport at $T=0$, in the superconducting state, 
should be similar to the $a$-$c$ anisotropy of transport in the normal state.
This is what is observed in the quasi-2D iron-based superconductor KFe$_2$As$_2$ \citep{Reid_2012_SUST,Reid_2012_PRL},
for example.
Explicitly,
$(\kappa_{\text{c0}}/T)/(\kappa_{\text{a0}}/T) \simeq (\kappa_{\text{cN}}/T)/(\kappa_{\text{aN}}/T)$, 
and we therefore expect
$\kappa_{\text{c0}}/T \simeq 0.2~\kappa_{\text{cN}}/T = 13 \pm 1 $\,\unitsmu,
since we have
$\kappa_{\text{a0}}/T = 0.2~\kappa_{\text{aN}}/T$~(Fig.~\ref{Fig1}) and
$\kappa_{\text{cN}}/T = 67 \pm 7$\,\unitsmu~
(see Fig.~\ref{Fig3}b).
This is in good agreement with the 
experimental value quoted above
($10 \pm 5~\mu$W~/~K$^2$~cm).
We conclude that the line nodes in the gap structure of \SRO~are vertical.

In most theoretical proposals, 
the gap minima do occur along vertical lines.
Presumably, some of these minima could accidentally be so deep as to produce nodes.
Let us consider different options.
%
%
First, a scenario of line nodes present only on the $\alpha$ surface is unrealistic because the full contribution 
of this small surface to the total in-plane conductivity
in the normal state is only 18\%~of $\kappa_{\text{aN}}/T$~\citep{Bergemann_2003_AP}, less than the zero-field fraction of 20\% (Fig.~\ref{Fig1}b).
In other words, the entire $\alpha$ Fermi surface would have to be normal already at $H=0$.
Such an extreme multi-band character is ruled out by 
two facts:
1) the residual linear term in $C_e/T$ at $T \to 0$ is too small \cite{Deguchi_2004_JPSJ};
2) an increase in impurity scattering
does not cause $\kappa_{\text{a0}}/T$ to decrease \cite{Suzuki_2002_PRL} --
unlike in CeCoIn$_5$, where electrons on part of the Fermi surface 
are uncondensed  and $\kappa_{\text{a0}}/T \propto 1 / \Gamma$ \cite{Tanatar_2005_PRL}.

Secondly, a scenario with nodes only on the $\beta$ surface is unlikely 
because the $\beta$ surface accounts for 80\% of the total normal-state conductivity 
along the $c$~axis, but only 37\% along the $a$ axis \citep{Bergemann_2003_AP}.
As a result, if only the $\beta$ surface had nodes, it would alone be responsible for 
the ratio 
$(\kappa_{\text{a0}}/T)/(\kappa_{\text{aN}}/T) = 0.2$,
and it would then necessarily produce a larger ratio along the $c$~axis (by a factor $\sim 80/37$),
giving
$(\kappa_{\text{c0}}/T)/(\kappa_{\text{cN}}/T) \simeq 0.2~(80/37) = 0.43$, so that
$\kappa_{\text{c0}}/T \simeq 29$\,\unitsmu.
Such a large value is not possible, since it exceeds the full measured conductivity at $T = 50$~mK (including phonons)~
(Fig.~\ref{Fig2}c).
Invoking line nodes on both $\alpha$ and $\beta$ surfaces decreases these estimates to 
$(\kappa_{\text{c0}}/T)/(\kappa_{\text{cN}}/T) \simeq 0.2~(89/55) = 0.32$
and $\kappa_{\text{c0}}/T \simeq 22$\,\unitsmu~-- 
still too large.
%

In summary, quantitative analysis indicates that the vertical line nodes in \SRO~are present
on more than one sheet, including the $\gamma$ sheet (\eg~on $\gamma$ and $\beta$), 
and most likely present on all three sheets of the Fermi surface.
%
This is consistent with the 
nodal structure
of a $d_{x^2-y^2}$~pairing state (with line nodes on all three sheets)
and that of a $d_{xy}$~state (with line nodes on $\gamma$ and $\beta$, but not $\alpha$).
The data would also be consistent with a $p$-wave pairing state with minima on $\gamma$ and $\beta$ that are so deep that they
extend to negative values and hence produce accidental nodes.


\section{FIELD DEPENDENCE}

Applying a magnetic field is a sensitive way to probe the low-lying excitations in a type-II superconductor \cite{Shakeripour2009}.
In the absence of nodes, the quasiparticle states are localized in the vortex cores, and heat conduction
proceeds by tunnelling between adjacent vortices, which depends exponentially on inter-vortex separation.
As a result, \Kzero~grows exponentially with $H$, as observed in all $s$-wave superconductors,
\eg~LiFeAs \cite{Tanatar2011}.
In a two-band $s$-wave superconductor like NbSe$_2$ \cite{Boaknin2003},
the exponential increase is seen below $H^\star <<$~\Hc, the effective critical field of the band with the minimum gap.  
By contrast,
in a nodal superconductor quasiparticle states are delocalized even at $T=0$ and $H=0$.
Increasing the field immediately increases their density of states, causing the specific heat to increase
as $\sqrt{H}$, the so-called Volovik effect.
As a result, \Kzero~grows rapidly with $H$ at the lowest fields \cite{Vekhter1999}, as observed in $d$-wave superconductors,
\eg~YBa$_2$Cu$_3$O$_y$ \cite{Chiao1999}.


\begin{figure}[t]
\centering
\includegraphics[width=7.8cm]{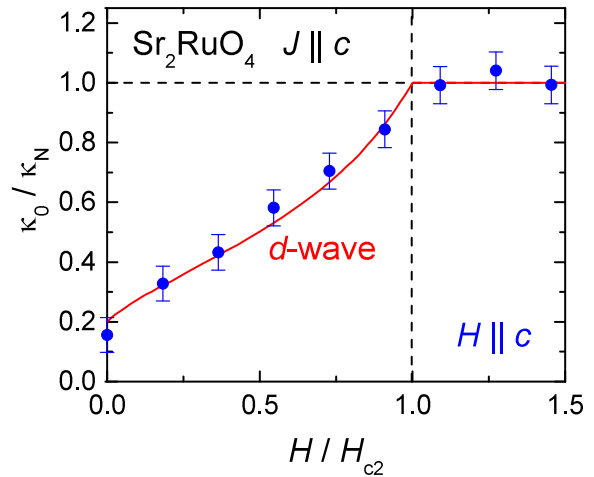}
\caption{
Residual linear term \Kzero~as a function of magnetic field, for a field along the $c$~axis ($H \parallel c$)
and a heat current along the $c$~axis ($J \parallel c$).
The data are plotted as (\Kzero)/(\KN)~$vs$ $H/$\Hc, with \Hc~$= 0.055$~T.
For this field direction (in the longitudinal configuration with a tiny \Hc), the magneto-resistance
in the normal state is negligible, and so \KN~is a constant below \Hc.
The data points are \Kzeroc~obtained by fitting \Kc$/T$~$vs$ $T$ as in Fig.~\ref{Fig2}b 
(red fit).
%
The solid red line is a theoretical calculation for a single-band $d$-wave superconductor \cite{Kusunose2002}.
}
\label{Fig4}
\end{figure}


In Fig.~\ref{Fig2}c,
we show $c$-axis data at $H = 25$~mT.
We see that even this tiny field (\Hc/50) induces a substantial increase in \Kc$/T$~at $T \to 0$.
This proves the existence of nodal quasiparticles with $c$-axis velocity.
In Fig.~\ref{Fig2}d,
a plot of \Kc$/T$~vs $H$~shows how rapid the rise is, whether 
\Kc$/T$~is measured at $T = 60$~mK (open circles) or extrapolated to $T=0$,
either linearly as in Fig.~\ref{Fig2}c (crosses) or through the fit described in sec.~IV (full red circles).

In Fig.~\ref{Fig3}, we show the $H$ dependence of \Kzero~in \SRO~($H \parallel a$), for both current directions.
Both \Kzeroa~and \Kzeroc~have the dependence expected of nodal superconductors,
as calculated for a single-band $d$-wave superconductor \cite{Vekhter1999}.
%
%

In Fig.~\ref{Fig3}c, we compare the $H$ dependence of \Kzeroa~and \Kzeroc~in normalized units, both plotted as 
(\Kzero)/(\KN)~$\equiv \kappa_0 / \kappa_{\rm N}$~$vs$ $H/$\Hc.
We obtain the normal-state conductivity \KN~below \Hc~by 
extending a fit of the data above \Hc~to lower fields.
For $J \parallel a$, there is negligible $H$ dependence up to 4~T, and so we take \KN~to be constant (Fig.~\ref{Fig3}a). 
For $J \parallel c$, \SRO~exhibits a sizable magneto-resistance, which varies as $H^2$ below 2~T or so \cite{Hussey1998}.
By the Wiedemann-Franz law, this implies that \KN~$= a / (b + c H^{2})$.
A fit of the data above \Hc~to this formula yields the red line in Fig.~\ref{Fig3}b.

The data in Fig.~\ref{Fig3}c are striking:
the two normalized curves are the same, at all fields, within error bars.
This is strong confirmation that line nodes are vertical. Indeed, horizontal line nodes would 
inevitably produce a qualitative difference between the two current directions,
roughly $d$-wave-like (rapid) for \Kzeroa~and $s$-wave-like (exponential) for \Kzeroc.
The fact that both curves in Fig.~\ref{Fig3}c are the same
 is also consistent with 
line nodes being present on all of the three Fermi surfaces.
Indeed, if line nodes were present only on the $\beta$ surface, for example,
\Kzeroc~would exhibit a $d$-wave-like $H$ dependence, as it is dominated by that surface,
while 
\Kzeroa~would exhibit an $s$-wave-like $H$ dependence, since it is dominated by the other Fermi surfaces.

The electronic specific heat at low temperature also displays a rapid increase at low field. 
In the $T=0$ limit, the residual linear term $\gamma_0(H)$ reaches $\sim 30$\% of its normal-state value
$\gamma_{\rm N}$ by $H \simeq 0.1$ \Hc~($H \parallel a$), and then increases more slowly at higher $H$
\citep{Nishizaki_2000_JPSJ,Deguchi_2004_JPSJ}.
We see from Fig.~\ref{Fig3}c, that the field dependences of $\kappa_0 / \kappa_{\rm N}$ and $\gamma_0 / \gamma_{\rm N}$ are similar.
To explain the rapid initial rise in $\gamma_0 / \gamma_{\rm N}$~$vs$ $H$, it was proposed that the
$\alpha$ and $\beta$ surfaces become normal at a field 
$H^\star \simeq 0.1$~\Hc~\citep{Deguchi_2004_PRL}.
But this is inconsistent with our data, since it would imply a much larger increase in $\kappa_0 / \kappa_{\rm N}$ for
$J \parallel c$ than for $J \parallel a$, given that the $\beta$ surface accounts for 80\%~of $\kappa_{\rm cN}/T$ but only
37\%~of $\kappa_{\rm aN}/T$.
This is not observed (Fig.~\ref{Fig3}c).

In Fig.~\ref{Fig4}, we show the effect of applying a magnetic field parallel to the $c$~axis.
This is the field direction for which the Volovik effect is the dominant excitation process,
and for which most theoretical calculations on quasi-2D superconductors have been carried out
(\eg~\cite{Vekhter1999}).
%
%
The overall field dependence of $\kappa_0 / \kappa_{\rm N}$ is in 
good agreement with calculations for a single-band 2D $d$-wave superconductor \cite{Kusunose2002},
as seen in Fig.~\ref{Fig4}.
%
Also, the specific heat of \SRO~exhibits a nice $\sqrt{H}$ dependence 
and detailed $\sqrt{H}/T$ scaling \cite{Deguchi_2004_JPSJ},
consistent with the behavior of a single-band $d$-wave superconductor \cite{Simon-Lee_1997_PRL}.


\section{SUMMARY}

In summary, our thermal conductivity measurements confirm that the gap structure of \SRO~has nodes
rather than deep gap minima and they reveal that those nodes are vertical lines along the $c$ axis.
%
%

In a nutshell, everything about the thermal conductivity of \SRO~is 
consistent with a $d$-wave
pairing state, including its absolute magnitude at $T \to 0$, 
its dependence on temperature, magnetic field and impurity scattering, 
and its isotropy relative to current direction.
%
A $d$-wave gap structure is also consistent with the specific heat of \SRO~\cite{Nishizaki_1999_JLTP,Nishizaki_2000_JPSJ,Deguchi_2004_PRL,Deguchi_2004_JPSJ},
including 
the magnitude of its jump at \Tc~and
its dependence on temperature, magnetic field
and impurity scattering.

Given that calculations find $p$-wave and $d$-wave solutions for \SRO~to be very close in energy \cite{Steppke2016},
it is tempting to consider a $d$-wave state for \SRO.
However, this comes into conflict with some important properties of the material, in particular
%
the absence of a drop in the NMR Knight shift below \Tc~\cite{Ishida_1998_Nature,Ishida_2016_PRB},
a signature of spin-triplet pairing,
and the onset of muon and Kerr signals below \Tc~\cite{Luke_1998_Nature,Xia_Kerr}, 
evidence that time-reversal symmetry is broken.
%
These are the natural properties of a chiral $p$-wave superconductor.
%
Note that the spin-singlet chiral $d$-wave state also breaks time-reversal symmetry, but its gap function varies as 
$k_z (k_x + i k_y)$ and therefore has symmetry-imposed line nodes
that are horizontal, not vertical \cite{Mazin_2005_PRL}.

We are therefore faced with a situation where \SRO~appears to adopt 
a $p$-wave state with a $d$-wave-like gap structure.
An intriguing solution to this conundrum has been proposed in the so-called $f$-wave state \cite{Hasegawa_2005_JPSJ,Graf_2000_PRB}, 
a combination of $B_g$ and $E_u$ representations ($B_g \times E_u$), where 
$B_g$ is either $B_{1g}$ ($d_{x^2-y^2}$) or $B_{2g}$ ($d_{xy}$),
with gap functions that vary either as $(k_x^2 - k_y^2)(k_x + i k_y)$ or as $(k_x k_y)(k_x + i k_y)$, respectively.

Further theoretical and experimental work is needed to resolve the puzzle presented to us
by the superconducting state of this exceptionally well characterized and otherwise rather conventional three-band metal.


%


\section{ACKNOWLEDGEMENTS}

We thank 
J.~Corbin, 
S.~Fortier, 
A.~Juneau-Fecteau, 
and
F. F.~Tafti 
for their assistance with the experiments,
and
A.~Balatsky,
M.~Graf,
A.~P.~Mackenzie,
K.~Samokhin,
M.~Sato, J.~Sauls,
T.~Scaffidi,
R.~Thomale,
and S.~Yonezawa
for stimulating discussions.
L.T. acknowledges support from the Canadian Institute for Advanced Research (CIFAR) 
and funding from the National Science and Engineering Research Council of Canada (NSERC), 
the Fonds de recherche du Qu\'ebec - Nature et Technologies (FRQNT), 
the Canada Foundation for Innovation (CFI) and a Canada Research Chair.
The work in Japan was supported by the JSPS KAKENHI (No.~JP15H05852).


\end{document}